# STRUCTURAL FEATURES OF SUN PHOTOSPHERE UNDER HIGH SPATIAL RESOLUTION

[1,2]A. A. Solov'ev, [1]L. D. Parfinenko, [1]V. I. Efremov, [1]E.A.Kirichek, [1]O.A.Korolkova

[1]Central (Pulkovo) Astronomical Observatory, Russian Academy of Sciences, St. Petersburg,

[2]Kalmyk State University, Elista, Russia.

*The fundamental small-scale structures such as granules, faculae, micropores that are observed in the solar photosphere under high resolution are discussed. As a separate constituent of the fine structure, a continuous grid of dark intergranular gaps or lanes is considered. The results due to image processing of micropores and facular knots obtained on modern adaptive optic telescopes are presented. For intergranular gaps and micropores, a steady-state magnetic diffusion mode is defined, in which horizontal-vertical plasma flows converging to a gap (and micropores) compensate for the dissipative spreading of the magnetic flux at a given scale. A theoretical estimate of the characteristic scales of these structures in the photosphere is obtained as 20–30 km for the thickness of dark intergranular spaces or lanes (and the diameter of the thinnest magnetic tube in the solar photosphere), 200–400 km for the diameter of micropores. A model of a facular knot with the darkening on the axis, which physically represents a micropore, stabilizing the entire magnetic configuration over a time interval of up to 1 day, is briefly described.*

## 1. Introduction

In photographs of the photosphere taken in the continuum, obtained with high angular resolution, the field of solar granulation appears as a complex, multi-scale structure (see Fig. 1). When observing at wavelengths λ ~ 3000Å, the usual photospheric granulation in the center of the disk begins to be replaced by facular granules. In the UV range around λ ~ 2000Å, conventional granulation is no longer visible. A two-dimensional picture of the power spectrum of the intensity of images obtained in the continuum reveals a large number of peaks (modes) that can noticeably change their frequency from place to place on the surface of the Sun. The most stable and pronounced scale of the fine structure of the photosphere is a granular grid of 1.5-2.4 arc seconds (1 "= 725 km). The amplitudes of higher-frequency harmonics are well below the 2σ confidence level and the corresponding formations are largely random (Muller, 1985). Note that the statistical method does not give the size of the structures themselves, but only the characteristic spatial size of the distribution grid of these structures. The transition to sizes of the structures themselves is possible considering how dense the data structures fit their plan form (i.e., what is the filling). For granulation the ratio is ~ 0.75 (Ikhsanov et al. 1997), so that the size of a typical photospheric granule is 1Mm (Fig.1).

The elements of the magnetic field's fine structure are closely related to the elements of the thin photospheric structure observed in white light (Rachkovsky et al., 2005). The problem of hidden magnetic fields invisible on the magnetograms was discussed in detail in the work (Stenflo, 2012). Small-scale magnetic elements are physically important from the point of view that they contain a large amount of free (associated with electric currents) energy, which can make a significant contribution both to the heating of the solar atmosphere and to the flare energy release.

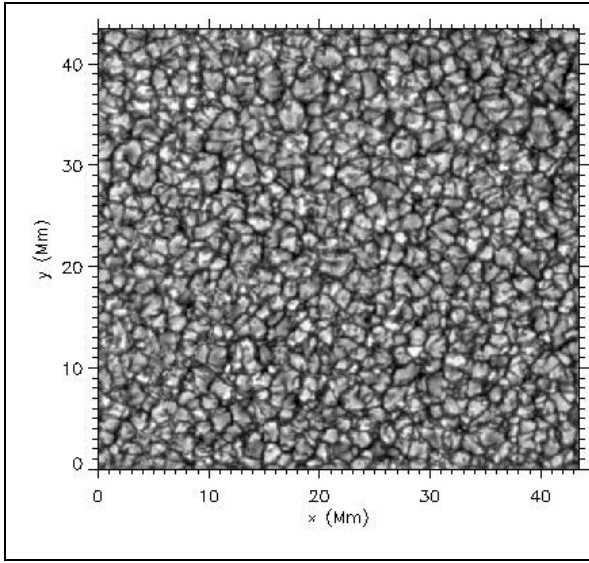

Fig. 1. Granulation image near the center of the solar disk, obtained on 30. 06.1970 at 8h44m10s (UT) at the Pulkovo stratospheric telescope (stratospheric balloon-borne observatory, Krat et al., 1970). Exposure time is 1 msec, flight altitude is 21 km. The filter selected wavelength band 4380-4800 A (effective $\lambda = 4600A$). An 80 mm FT30 film calibrated with a 9-step attenuator was used, minimizing the distortion of the granulation contrast. An angular resolution of 0".24, corresponding to a diffraction resolution of a 50 cm telescope lens, is achieved.

In 1973, an improved version of the stratospheric telescope with a mirror diameter of 100 cm was launched. Then, in several photos of the photosphere, the diffraction resolution of the lens was obtained as - 0".12 (Krat, 1974), which is considered to be a very good result even by today's standards. A photoelectric image quality analyzer that evaluated the sharpness of the granulation was used. Photographs and spectra of the Sun taken at the Pulkovo Stratospheric Telescope in the visible range remained to be one among the best for a long time. The 100 cm-German Spanish stratospheric solar telescope (Sunrise telescope) in the UV range that was launched in June 2009 possessed the best performance and has been equipped with a vector magnetograph (Barthol et al, 2010). The main advantage of the balloon launched stratospheric observatories over space borne observatories lies in the comparatively very cheap cost of the prior relative to the later respectively.

Over the past decade, the use of new technologies has led to a revolution in the quality of the resulting terrestrial images of the sun. The main ones are: a) adaptive optics, for example, Rimmele & Marino, (2011); b) multi-frame deconvolution (multi-object multi-frame blind deconvolution, MOMFBD), van Noort et al., (2005); c) speckle reconstruction (Wöger et al., 2008). Currently, several ground-based solar telescopes equipped with adaptive optics systems allow systematic high-resolution observations. They are equipped with large light mirrors made using modern technologies. Such are, for example, the 1.6 meter GST reflector (http://www.bbso.njit.edu) equipped with adaptive optics and image speckle reconstruction, the New Vacuum Solar Telescope (NVST) in China (http://fso.ynao.ac .cn / index.aspx) and others. These ground-based telescopes provide spatial resolution better than that achieved from the stratosphere and even from space for a given cost of the telescope. Powerful spectrographs allow you to explore the fine structure of magnetic fields. Ground-based observations are complemented by cosmic ultraviolet observations.

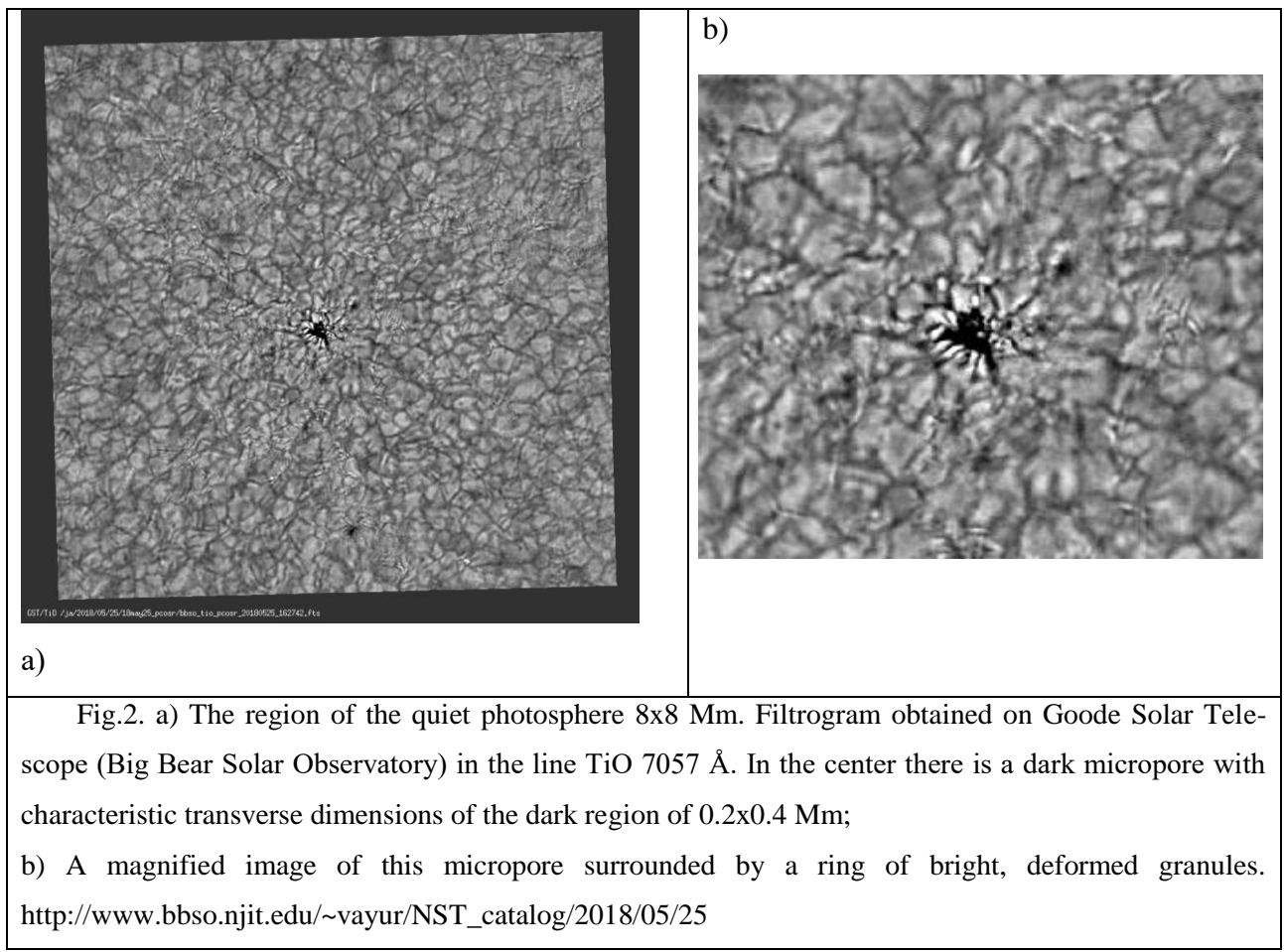

Fig.2. a) The region of the quiet photosphere 8x8 Mm. Filtrogram obtained on Goode Solar Telescope (Big Bear Solar Observatory) in the line TiO 7057 Å. In the center there is a dark micropore with characteristic transverse dimensions of the dark region of 0.2x0.4 Mm;

b) A magnified image of this micropore surrounded by a ring of bright, deformed granules. http://www.bbso.njit.edu/~vayur/NST_catalog/2018/05/25

In this paper, we intend to discuss the basic physical properties of the elements of the photospheric fine structure, estimate the minimum scales at which the condition for the magnetic field lines frozen in the plasma of these elements begins to break down and reveal the role of solar micropores in the formation and stabilization of facular nodes.

## 2. The fundamental elements of the photospheric fine structure.

In a quiet solar photosphere, the main structural element obviously is granulation, which has a characteristic spatial scale of about 1 Mm and a lifetime of 5 to 15 minutes. The photos-pheric granule is a characteristic manifestation of "overturning" convection, supplemented by a variety of wave processes, in which the hotter substance (and magnetic field) from the central part of the convective cell is transferred during a overturn to its cooling periphery. Here it is necessary to note the special role of descending currents, arising from the fact that in the layer where solar granulation is observed, radiative transfer begins to play a larger role, and a transition from almost adiabatic convection to a layer with free currents that descend to the periphery occurs in the form of jets (Rast, 2003).

A special small-scale structure of the photosphere as a consequence overturning convection appears as a grid of dark intergranular spaces (the net of dark intergranular lanes, for brevity - DIL). It represents a single, continuous physical formation, changing its geometric configuration with the same characteristic time as the pattern of the granular pattern, but never fading. The transverse size of the DIL is less than 0".3. In the active regions, the contrast is greater and the average width of the gaps or lanes, estimated at the level of the current resolution, is 0".1 (ie, 70 km) and even less (Schlichenmaier et al, 2016) . It is important to emphasize that intergranular spaces are characterized not only by a sharply reduced gas pressure due to a drop in temperature, but also by a rather strong kilogauss magnetic field, which provides a transverse pressure balance. They also observe descending laminar plasma flows, leading to an increase and ordering in the structure of the vertical magnetic field in these regions.

The emergence of magnetic fields with strengths greater than 150–200 Gs result in the appearance of facular fields in the photosphere which are weakly distinguishable on the disk, but clearly visible at the edge of the solar disk as they approach the limb. The facular fields usually appear in the active area before the spots, and disappear later, but each of the small faculae field represents a rather ephemeral formation with a lifetime of not more than 1 hour. In general, it is considered that there are three main types of magnetic structures in facular fields of active regions (Title et al.,1992; De Pontieu et al., 2006; Berger et al., 2004; Berger et al, 2007; Dunn & Zirker, 1973; Lites et al., 2004; Narayan & Scharmer, 2010; Kobel et al., 2011; Viticchi´e et al., 2011; Stenflo, 1973, 2017; Solanki, 1993):

1. Short-lived small faculae of granular scales with a lifetime of 5-10-15 minutes and size - 0.5-1 angular second. The magnetic field strength in these elements is close to the condition of equidistribution of the energy density: $B^2(8\pi)^{-1} \cong 0.5\rho_{ph} \langle V_{turb}^2 \rangle$, где $\rho_{ph} \cong 2\times 10^{-7} g/cm^3$ - photospheric plasma density, a $V_{turb} \cong 10^5 cm/s$ - characteristic speed of turbulent pulsations. Note that, according to the given numerical values, the dynamic pressure of turbulent motions $0.5\rho_{ph} \langle V_{turb}^2 \rangle$ is only $10^3 dyn/cm^2$, which is two orders of magnitude less than the gas pressure at the level of the photosphere ($P_{ph}$ ; $1.5\times 10^5 dyn/cm^2$).With such a dynamic pressure in the photosphere, the condition of equidistribution of energies is fulfilled for *B*(*equipartition*) = 150-200G.

Facular granules are very dynamic and are in constant motion due to the effects on them from the fields of granulation and supergranulation. Their brightness is relatively small and it is related apparently due to the interaction with a magnetic field whose pressure is close to the dynamic pressure of convection. As a result the granules are tightened and consequently structured by this magnetic field leading in their increased brightness that is seen when the granules are overturned with their lateral surfaces exposing through the relatively transparent magnetic flux tubes. In general, the physical nature of these elements is well described numerically within the framework of the concepts of magnetoconvection (Keller et al., 2004; De Ponteieu et al, 2006; Berger et al, 2007).

2. One observes facular nodes (facular knots), relatively stable large-scale magnetic formations with a lifetime of one hour or more i.e. up to 1 day and with sizes up to several Mm and a magnetic field strength of 300 to 1200 Gs against the background of small facular granules. Facular nodes have a central depression (in fact, they contain a micropore stabilizing them in their center) and due to this they are similar in their properties to solar pores. Apparently, these objects are located at the points where several convective supergranulation cells merge. As is known in these cells radial-horizontal plasma flows are observed which concentrate several dozen magnetic facular elements having the form of individual magnetic flux tubes or bundles in inter-supergranular gaps. This rakes them due to the frozen field in the plasma to the edges of the cells. These currents and most importantly the gas pressure in the intergranular gaps which is significantly lower compared to the photospheric gas provide a sufficiently high radial magnetic field strength leading to a steady long-term existence of facular assemblies.

3. Pores are small spots without penumbra with lifetime of several days and lateral dimensions of 2-4 Mm and sometimes ranging up to 8 Mm. They have magnetic field strengths of about 1500 Gs. Of particular importance is the presence of a large number of micropores in the facular areas. Micropores are formations that differ from the pores only in that they have smaller sizes i.e. less than 1 Mm, but have the same high contrast with sharp photometric boundaries and a lifetime from several hours to 1 day..

One can classify polar faculae into a separate group as they are found at latitudes above 70 degrees with lifetimes ranging from several hours to days. Their magnetic field strength is about 1 kG (Homann et al., (1997); Okunev & Kneer (2004); Blanco Rodr´ıguez et al., 2007). These objects are remarkable in that they reflect the solar magnetic cycle in their development.

The maxima of their appearance is shifted by 5–6 years relative to the sunspot cycle, i.e. they anti-correlate with the sunspot cycle (Makarov & Sivaraman, 1989; Makarov et al., 2003). In this paper, we will not deal separately with these structures..

The most famous model of the solar faculae is the so called model of "hot walls" (Spruit, 1976). According to this model (Knoelker et al., 1988; Grossmann-Doerth et al., 1994; Topka et al., 1997; Okunev & Kneer, 2005; Steiner, 2005), faculae are indentations in the photosphere created by magnetic flux tubes. The walls of these tubes are hot, and the temperature at the bottom of these tubes depends on the diameter of the tube. The bottom is cold if their diameter is more than 300 km, and hotter if it is <300 km. Due to the magnetic pressure, the temperature in the thick tube is lower than in the surrounding atmosphere at the same geometric height. Therefore, the bottom of the tube becomes dark in the observations. But if the tube is narrow enough, it can be heated by horizontal radiation transfer and becomes noticeably brighter. The contrast of the tube depends not only on its diameter, but also on the strength of the magnetic field of the tube: the larger the field, the darker the bottom of the tube. Proponents of this model believe that signatures of these hot granular walls ("faculae") can be seen (Lites et al. (2004); Kobel et al. (2009); Hirzberger & Wiehr (2005); Berger et al. (2007). The hot wall model describes well the situation with the appearance of faculae on the solar disk, but fails completely when going to the limb: Libbrecht & Kuhn (1984); Shatten et al., (1986); Wang & Zirin, (1987). The problem is that according to this model, torches on the limb should not be observed at all, because in this case the line of sight of the observer is perpendicular to the vertical axis of the magnetic tube, and none of its deep and hot walls cannot be observed for a purely geometrical reason. Meanwhile, as is well known, it is on the limb that the faculae are best seen (see, for example, Wang & Zirin, 1987). This fundamental contradiction clearly shows the inconsistency of the hot wall model. An alternative model for solar faculae in the form of a hot vertical magnetic tube with a horizontal substrate is described in (Solov'ev & Kirichek, 2019). A brief description of the flare assembly model is given below in section 4.

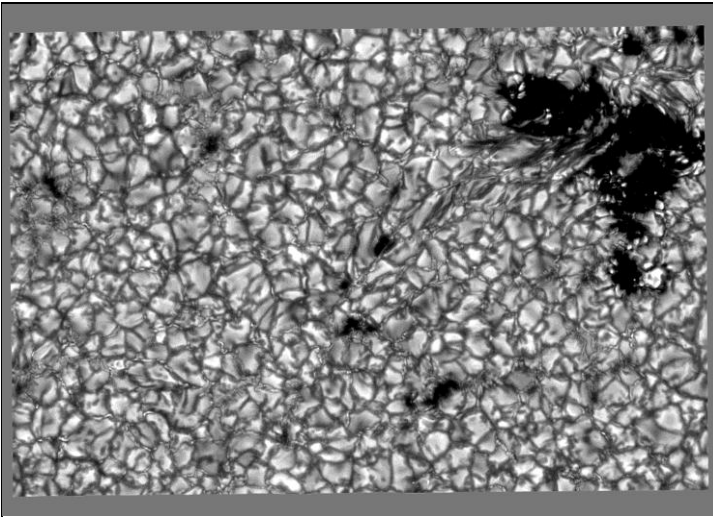

Fig.3. In the upper right corner there is a large pore in which the formation of penumbra is already noticeable and smaller dark formations scattered across the field of the picture are micropores with high contrast. They have diameters of about 400 km and a lifetime of many hours.
Image taken 2016-09-19, 9h 28m 36s, SST telescope.

The criterion of high spatial resolution underwent significant changes after the appearance of new optical solar ground-based telescopes. High angular resolution is currently determined by the visualization of structures on the Sun with a size of less than 180 km or in angular units less than 0".25. The Vacuum Tower Telescope (VTT) which uses a combination of adaptive optics (AO) and speckle interferometric image reconstruction, is capable of achieving a spatial resolution of 0".18 in the photospheric continuum (500 nm). The resolution of Dunn Solar Telescope (DST) in the G 430.5 nm passband is 0".14 or about 100 km on the Sun ( Berger & Title, 2004). The diffraction limits on the resolution of the Swedish 1-meter Solar Telescope (SST) in the G-band and in the 393.3 nm line of Ca IIK are 0 ".11 and 0" .10 respectively. Both of these limits were reached using the 37-element AO system (Scharmer, 2003). The New Solar Telescope of 1.6 m aperture at the Big Bear Solar Observatory at a wavelength of 705.68 nm reached the diffraction limit of the lens which is 0".09 (Andic, 2013). By using the best speckle reconstructions at 589nm wavelength at 1.5m GREGOR telescope at the Observatorio del Teide in Tenerife, Spain (Schmidt et al. 2012), a resolution of 0".08 was obtained which corresponds to 60 km on the Sun (Schlichenmaier et al., 2016). In order to check how accurately the contrast of details is preserved during speckle reconstructions of photospheric photos, a simultaneous comparision of the images taken by the Hinode and those obtained from speckle reconstructions of one and the same object with the DST telescope of the National Solar Observatory is carried out and as result a good agreement between these two images is observed (Wöger et al., 2008).

As for the time resolution, the current standard for it is a cadence of less than 10 seconds (The Dutch Open Telescope).

Thus today the spatial and temporal resolution of images obtained for example by the SDO space observatory (~ 1 ") is already lagging behind the resolution of ground-based instruments and does not allow to solve some problems concerned with ultrafine structure. With in-creasing diameter of ground-based telescopes and further improvement in image processing techniques give marked advantage to the ground based observatories over the space based ones. But the advantage of space telescopes is that they allow us to systematically receive long continuous series of observations in inaccessible terrestrial electromagnetic spectrum. Thus, the space and ground observations are complementary with respect to each other and their advantages and limitations are determined by the nature of ongoing research.

Small-scale structures show up particularly well in filtergrams in the photosphere at TiO 7057 Å (bandwidth: 10 Å) (Titanium (II) oxide (TiO) spectral line, dark-red) and in the G band (430.5 ± 0.25 nm (G-band, blue-ish). Due to high transmission bandwidth of the filter for the TiO 7057 Å line, the image on the resulting filtergrams does not depend much on the characteristics of the line itself and one observes structures almost at the photospheric level $\tau 500 = 1$ (Sinha & Tripathi, 1991). In this spectral line bright microstructures appear smaller than the resolution achieved (Andic et al., 2011), and they probably have magnetic flux tubes with field strengths of about 1 kG. In the quiet photosphere the 7057 Å TiO spectral line is very weak. Figure 4 shows the filtergram of the quiet photosphere in the TiO 7057 Å line, obtained on BBSO on June 19, 2017 at 17h 00m 05s. The two dark structures in the center of the image are typical micropores..

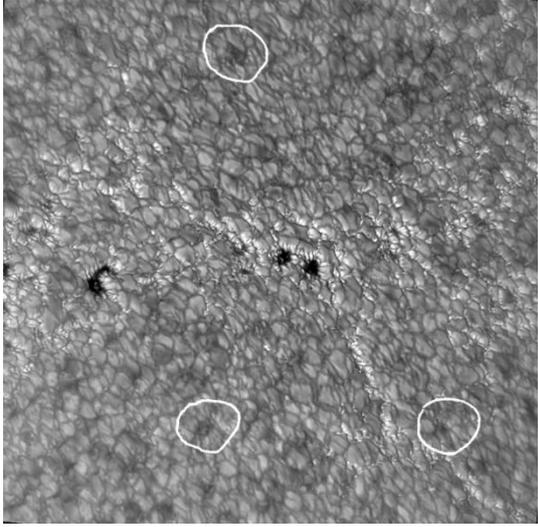 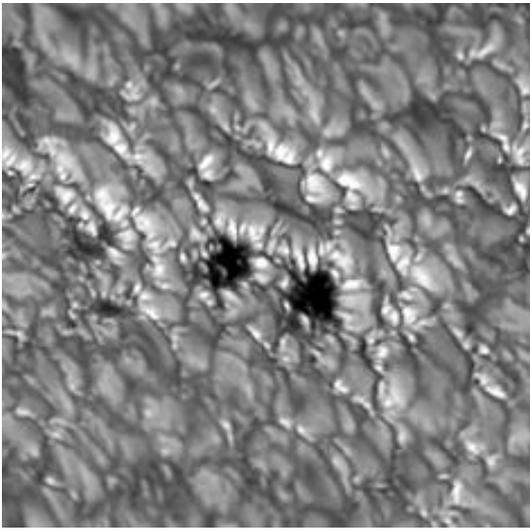

| Fig.4a. Adjacent extended areas representing a strong magnetic field are visible near the micropores, where the granulation pattern is small-scaled and irregular (anomalous granulation, Dunn & Zirker, 1973). Micropores showing tendency of intensity darkening and | Fig.4b. Two central micropores, which, like the micropores in Fig. 2, have a diameter of about 0.3–0.4 Mm and also surrounded by bright annular and semi-annular rings, which is a characteristic property in the facular regions. |

| showing strong flux concentrations are marked. According to Narayan & Scharmer (2010), they are called micro-protopores. | |

| 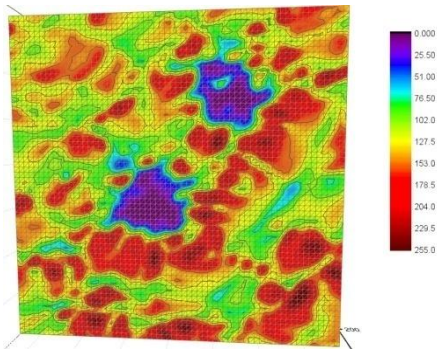 | 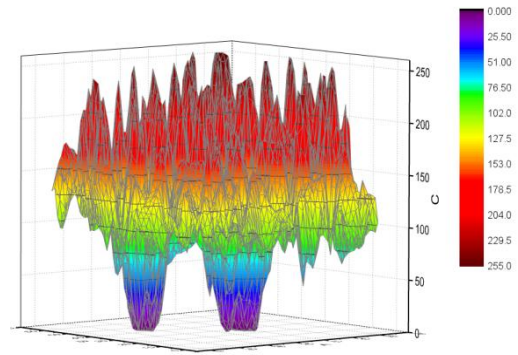 |
| Fig.5a. Site with micropores in fig. 4b, depicted in pseudo colors. From the bottom and to the right, the micropores are surrounded by bright patches, divided in azimuth into individual elements. Left and top bright elements are less common. | Fig.5b. Three-dimensional section of the same micropores. The spatial coordinates are plotted along the X and Y axes, and the brightness at the corresponding points of the micropores along the Z axis. |

## 3. Estimating the minimum size scale of the solar photospheric magnetic structures.

As a result of significant advancements and improvements in the techniques of recent photospheric and chromospheric optical imaging, the question arises as to how close are the modern telescopes' resolution in resolving the smallest possible quasi stable magnetic structures on the photosphere and what should be the size estimate of such a small object. Recall that the New Solar Telescope with 1.6 m aperture at the Big Bear Solar Observatory working at a wavelength of 705.68 nm reaches the diffraction limit of its lens which is 0".09 (Andic, 2013). This corresponds to a size of ~ 65km on the surface of the Sun. According to ( Stenflo, 2012) about 80% of the total flux of magnetic structures are invisible even in Hinode resolution. The photosphere is a swirling ocean of intricate magnetic fields and a tool like Hinode can only see "the tips of icebergs". Size of "self magnetic elements" is estimated to be 120 times less than the resolution of Hinode (about 250 km), i.e. it gives only 2 km (Stenflo, 2017).

### 3.1. Diffusion process in a narrow intergranular gap.

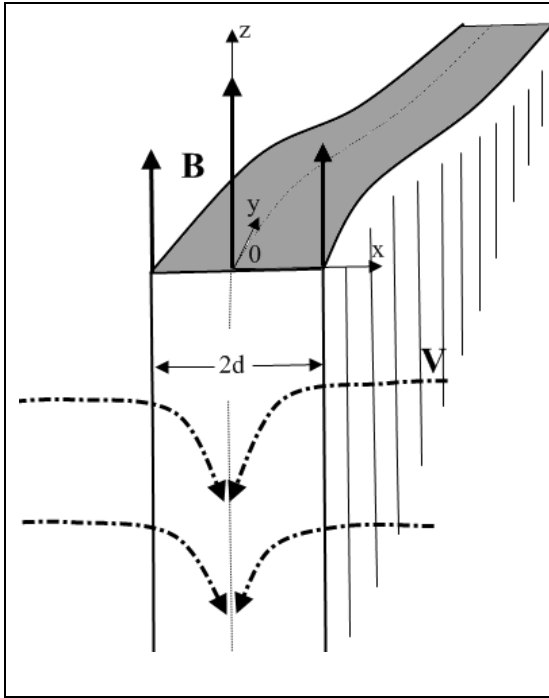

Fig.6. DIL is presented as a vertical magnetic layer of thickness 2d. The magnetic field B has only a vertical component, the velocity field V is directed to the layer and presses the diffusing magnetic field to the center of the layer.

We will analyze the processes of magnetic diffusion in DIL and in micropores, based on the equation for induction:

$$\frac{\partial \mathbf{B}}{\partial t} = -\eta \cdot rot(rot\mathbf{B}) + rot[\mathbf{V} \times \mathbf{B}], \qquad (1)$$

where $\eta = \dfrac{c^2}{4\pi\sigma}$ - plasma magnetic viscosity with normal gas-kinetic conductivity $\sigma$, and $\mathbf{V}$ – plasma flow rate in the photosphere. Imagine DIL as a vertical magnetic wall of width 2d (Fig. 6), pressed from both sides by horizontally vertical laminar plasma flows converging to the center of the layer, which are formed by the surrounding granules. We assume that the changes of any parameter along the y and z coordinates are much smaller than the transverse one: $\left|\dfrac{\partial}{\partial z}\right|, \left|\dfrac{\partial}{\partial y}\right| << \left|\dfrac{\partial}{\partial x}\right|$. The magnetic field in the DIL is vertical and distributed in the transverse direction according to the law:

$$B(x,t) \equiv B_z(x,t) = B_0 \exp(-k^2 x^2), \qquad (2)$$

where $B_0$ - magnetic field strength at the middle line DIL, и $k = d^{-1}$ - inverse half layer thickness. The velocity of laminar flows accumulating magnetic field in a DIL can be represented as: $\mathbf{V} = \{V_x(x)\mathbf{e_x}, 0, V_z(x)\mathbf{e_z}\}$, where:

$$V_x = -\gamma x \mathbf{e_x}, \quad V_z = -\frac{1}{1+\gamma x}\mathbf{e_z}. \qquad (3)$$

They describe currents converging to the middle line of the magnetic layer. Here $\gamma = t_{dyn}^{-1}$ - inverse time, characterizing the dynamics of currents in the granule; $t_{dyn}$ – «dynamic» time, in fact, this is the time of one complete cycle of the convective cell - granules. Substituting expressions (2), (3) into equation (1), we get:

$$\frac{\partial B}{\partial t} = B\left[-2\eta k^2 + 4\eta k^4 x^2 + \gamma - 2\gamma k^2 x^2\right] = -B\left[(2\eta k^2 - \gamma)(1 - 2k^2 x^2)\right]. \quad (4)$$

This shows that there exists such a layer thickness $k_0 = d_0^{-1}$ at which the right-hand side of equation (4) vanishes, i.e. the system comes to a stationary state and the field ceases to change with time. At this scale the diffusion of the magnetic field characterized by time $t_{dif} = (2\eta k_0^2)^{-1}$ is exactly compensated by the increase in the magnetic field strength due to its dynamic sweeping to the central line of DIL:

$$2\eta k_0^2 - \gamma = 0. \quad (5)$$

At $x < d_0$ magnetic diffusion dominates in the system and at $x > d_0$ the effect of magnetic flux freezing prevails. Relation (4) allows us to estimate this characteristic scale of the stationary state numerically as $d_0 = \frac{c}{\sqrt{2\pi\sigma\gamma_{gr}}}$. Therefore one obtains for the dynamic time of a granule $\gamma_{gr}$ as

$$\gamma_{gr} = \frac{V_{gr}}{L_{gr}}; \quad \frac{10^5 \, cm/s}{10^8 \, cm} = 10^{-3} \, s^{-1}, \quad (6)$$

where $V_{gr}$ is the velocity of currents in the photospheric granule (about 1 km / s), and $L_{gr}$ - characteristic granule scale (about 1 Мм). To assess the plasma conductivity in the DIL region, one can draw the following considerations: the conductivity of the photospheric plasma at T = 5800 K is approximately $\sigma_{ph}$ ; $10^{12} \, s^{-1}$, and the conductivity of the plasma in the shade of a sunspot at a temperature of about 3700-3800 K is equal to $\sigma_{sp}$ ; $10^{10} \, s^{-1}$ (Obridko, 1985). According to its characteristics, the DIL plasma probably occupies an intermediate state between the plasma of the sunspot shadow and the photospheric plasma, therefore a reasonable estimate for it would be $\sigma_{DIL}$ ; $10^{11} \, s^{-1}$. Substituting the above numerical values into the formula for $d_0$, we get $d_0(DIL) = 12$ км. Consequently, the characteristic thickness of a stationary DIL will be $2d_0(DIL) \approx 24 km$. This size is lower than the current resolution of the best ground-based telescopes, but it fits well in the range of 10-100 km, in which, according to Stenflo (2012), there is a peak of the size distribution of magnetic fields that are invisible on the magnetograms, existing in the form of thin magnetic flux tubes.

### 3.2. Diffusion of the magnetic field in the micropore.

Let us turn to the analysis of diffusion of magnetic field in a micropore. In this case, we use a cylindrical coordinate system and the vertical magnetic field profile is taken as:

$$B_z(r,\varphi,t) = B_0 \exp(-(kr^2)) + B_0 \sum_i b_i (kr)^{n_i} \sin(m_i\varphi + \varphi_{0,i}) \exp(-a_i(kr^2)), \quad (7)$$

where the first term corresponds to the main axisymmetric mode, and the angle-dependent terms under the sum sign describe the field deviations from axial symmetry (as can be seen from Figures 2–5, the perimeter of the pore is usually strongly cut, therefore, taking account of azimuthal field variations is necessary). Parameters $a_i, b_i, n_i, m_i, \varphi_{0,i}$ – are constants described by the azimuthal and radial inhomogeneties of the magnetic field. An example of the asymmetric form of the magnetic field in a micropore, described by the distribution of the form (7), is shown in Fig.7.

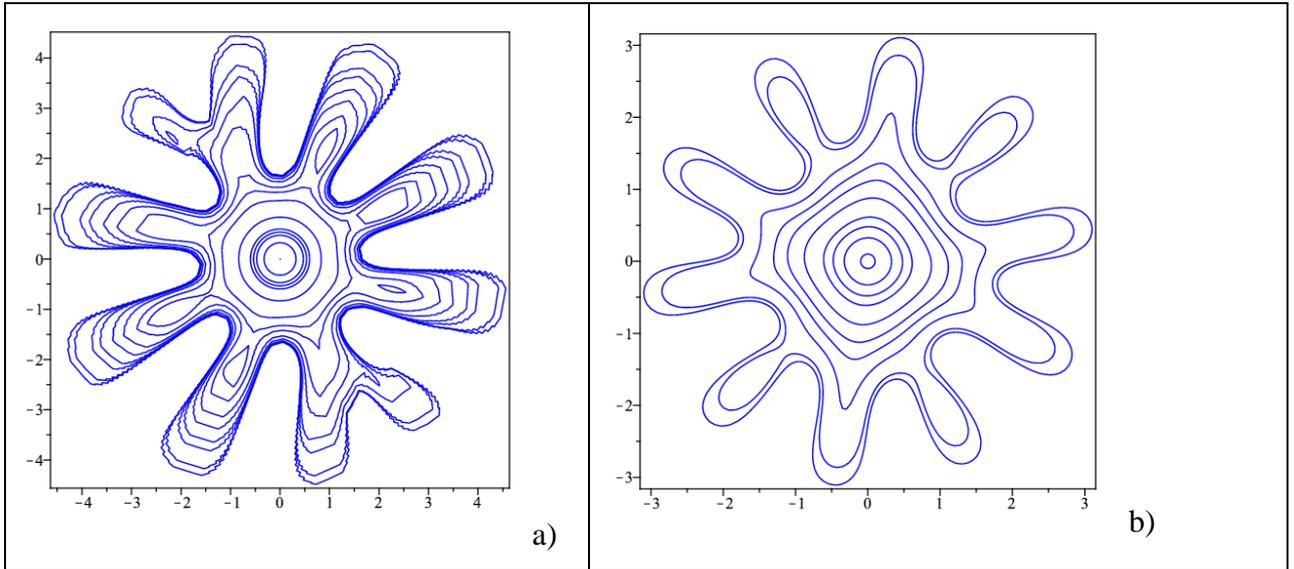

Fig.7. Examples of the shape of the magnetic field of a micropore (isogaussian) projected on a horizontal plane. a) the sum of three modes: main ($m_0 = 0$) and two asymmetric $m_1 = 8, m_2 = 10$:

$$B_z(r,\varphi,t) = B_0 \exp(-(kr)^2) + B_0 (0.01(kr)^8 \sin(8\varphi - 2) + 0.001(kr)^{10} \sin(10\varphi)) \times \exp(-0.75(kr)^2).$$

b). Superposition of modes: $m_0 = 0, m_1 = 4, m_2 = 10$. The unit distance is 0.1 Мм along the axes.

Substituting expression (7) into the induction equation and grouping the corresponding terms, we find for the main axisymmetric mode:

$$\frac{\partial \ln\left[B_0 \exp(-k^2 r^2)\right]}{\partial t} = -2(2\eta k^2 - \gamma_{mp})(1 - k^2 r^2). \qquad (8)$$

Here again, as a condition for a stationary state, we obtain an equation similar to (5):

$$2\eta k_0^2 - \gamma_{mp} = 0, \qquad (9)$$

but in this case it makes sense to reverse the problem: while analyzing DIL we looked for an estimate of such an unobservable parameter as the thickness of the intergranular gap $2d_0$, based on the average lifetime of a photospheric granule ($\approx 10^3 c$), while in the case of micropores it is evident that the spatial width is about $2k_0^{-1} = 2r_0$ ; $400 km$ from the modern high resolution observations.

Considering this estimate as the basis and imposing the condition for plasma transperancy of the micropores close to the umbra of the sunspot ($\sigma_{mp} = 3 \times \sigma_{sp} = 3 \times 10^{10} s^{-1}$), we can estimate dynamic time $\gamma_{mp}$ from the relation (9), thus obtaining the characteristic lifetime of a solar micropore surrounded as shown in the picture by tens of deformed photospheric granules: $\gamma_{mp}^{-1} = \tau_{mp} \approx 10^5$ s $\approx 1 day$. This assessment is generally consistent with the observations (Anđić, 2013) and the fact that the micropore in the center of the facular assembly stabilizes this magnetic formation within a time interval of up to 24 hours..

Since in this case we are considering a cylindrical magnetic configuration, it is appropriate to ask the question as to what is the diameter of the thinnest magnetic tube in the solar photosphere? If in formula (9) we take the same dynamic time as for the intergranular lane or gap i.e. $10^3$ sec and the same conductivity, $\sigma = 10^{11} s^{-1}$, then by virtue of the fact that formulae (5) and (9) coincide in form, we find that the diameter of the thinnest magnetic flux tube in the solar photosphere is 25 km. This corresponds to an estimate of 15–30 km obtained independently by Botygina et al (2016; 2017) when analyzing HINODE / SOT data.

Inorder to clarify the uncertainty in our estimate of the gas kinetic contuctivity of the plasma $\sigma$, we depict in Fig. 8 the dependence of the thickness (width) of a magnetic structure on its lifetime obtained from equation (9) for two different values of conductivity. As conductivity $\sigma$ happens to be under the radical sign in the formal for the skin time, its variation doesn't strongly influence the result which is justified from the fact that the diameter of micropores with a dynamic time of about a day lies within 240-440 km.

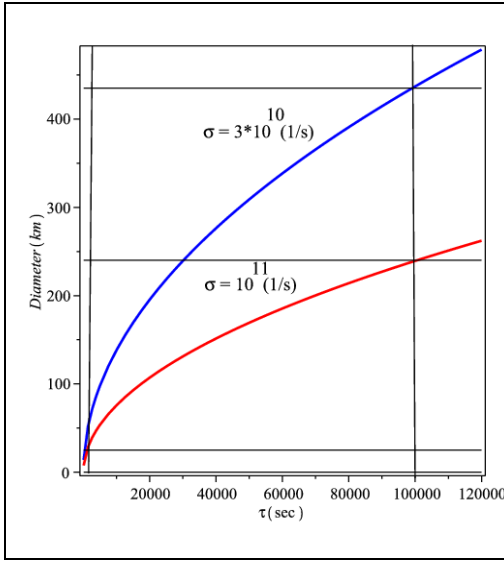

Fig.8. Dependence of the width of a magnetic element on its lifetime for two different values of plasma conductivity: $\sigma = 3\times 10^{10}(1/s)$ ( blue curve) and $\sigma = 10^{11}(1/s)$ ( red curve). Thin horizontal lines near the bottom and to the top mark the level where the diameter is 24 km (the thickness of the intergranular gap or the diameter of the thinnest magnetic tube with dynamic time $\tau_{gr} = 10^3 s$) and 440 km — the diameter of the micropore with. $\tau_{mp} = 10^5 s$ respectively.

For an asymmetric mode of the form (7) with $m_i \neq 0$ we have:

$$\frac{\partial \ln\left[ B_0 b_i (kr)^{n_i} \sin(m_i\varphi + \varphi_{0,i})\exp(-a_i(kr)^2)\right]}{\partial t} = $$
$$= -(n_i+1)\left[\eta \frac{m_i^2 - n_i^2}{(n_i+1)r^2} + 4\eta a_i k^2 (1 - \frac{a_i(kr)^2}{n_i+1}) - \gamma_i \frac{n_i+2}{n_i+1}(1 - \frac{2a_i(kr)^2}{n_i+2})\right]. \quad (10)$$

The exact solution describing the stationary state of a small-scaled fine structure of the field is almost impossible to obtain, but one can derive an approximate solution. First of all, we must set $m_i^2 - n_i^2 = 0$, to avoid divergence in the first term in the square bracket. Further, we note that for sufficiently large $n_i$ and small $a_i$ expressions $(1 - \frac{a_i(kr)^2}{n_i+1})$ и $(1 - \frac{2a_i(kr)^2}{n_i+2})$ in the square brackets differ little form each other, and then the approximate stationarity condition takes the form:

$$4\eta a_i k^2 - \gamma_i \; ; \; 0. \quad (11)$$

Here again it makes sense to consider the inverse problem i.e. by knowing the characteristic spatial scale of the fine structure elements (about 100 km, according to Fig. 7), one can estimate their time from formula (11):

$$\frac{a_i c^2}{\pi \sigma d_i^2} \; ; \; \frac{1}{\tau_i}. \quad (12)$$

Considering $a_i = 0.5, \; \sigma = 3*10^{10}(1/s), \; d_i = 10^7 cm$, we get $\tau_i \approx 2\times 10^4 c$. Thus, the fine structure of the magnetic field of the type shown in Fig. 7 is eroded due to ohmic diffusion during the characteristic time of 5-6 hours and then restored back in a different form over the same time as a result of micropore's accidental exposure to the surrounding photospheric granule.

**4. About the model describing the facular knots.**

In the work (Solov'ev & Kirichek, 2019) a model for a stationary facular knot was constructed in the form of a vertical, untwisted (no azimuthal field) magnetic tube, where the radial and vertical magnetic field components are dependant on all three variables of the cylindrical coordinate system ( $r, \varphi, z$ ) and are alternating in the radial and also in the azimuthal direction:

$$B_z(r,\varphi,z) = B_0 F(A,\varphi) b_z(r,z); \quad b_z = Z(kz) J_0(kr),$$
$$B_r(r,\varphi,z) = B_0 F(A,\varphi) b_r(r,z); \quad b_r(r,z) = [-\frac{\partial Z(kz)}{\partial (kz)}] J_1(kr). \qquad (13)$$

Here $B_0 = const$ – unit of measurement of magnetic field strength, $k$ - inverse spatial scale, $J_0(kr)$, $J_1(kr)$ - Bessel functions of zero and first order, $A(r,z) = \int_0^r b_z r dr$ - flow function defining the field components: $b_z(r,z) = \frac{1}{r}\frac{\partial A}{\partial r}; \; b_r(r,z) = -\frac{1}{r}\frac{\partial A}{\partial z}$. In this case: $A(r,z) = kr J_1(kr) Z(z)$. $F(A,\varphi)$ is an arbitrary function chosen dependant on the angle of rotation and magnetic flux A. The magnetic field (13) satisfies the condition $div\mathbf{B} = 0$. The function $Z(kz)$ which defines the height dependence of the field was chosen as:

$$Z(kz) = \frac{2}{\exp(kz)+1}. \qquad (14)$$

At the level of the photosphere, when $z = 0$, this function is equal to 1 and $B_0$ in the formula (13) is the field in the center of the configuration at the level of the photosphere. When $z > 0$ then this function tends to 2exp(-kz) which makes the field approximately potential. When $z < 0$ the magnetic field tends to $2B_0 = const$.

The dependence of the arbitrary function F on the angle was chosen as follows:

$$F^2(A,\varphi) = 1 + f(A,\varphi) = 1 + k^2 \left| A \cdot \sum_i a_i \sin(m_i \varphi) \right|, \qquad (15)$$

where $f(A,\varphi)$ is a positively oscillating function of $\varphi$ and diminishing with amplitude down the height as a result of the decreasing flux $A$. In (15) $a_i, m_i$ - are certain numerical coefficients, $k$ - inverse spatial scale. By choosing different values of the angular parameter $m$, one can obtain various forms of angular variations in the temperature profiles of faculae. It must be noted that when there is no angular dependence of the field we have $F = 1$. As for plasma flows described by this model in the stationary configuration under consideration, the Alfven Mach number which expresses the velocity of the currents in terms of the Alfven velocity $M_A = \dfrac{V}{V_A}$ is related to $f(A,\varphi)$ by the ratio:

$$M_A^2 = 1 - \frac{1}{F^2} = \frac{f}{1+f} > 0. \qquad (16)$$

The external environment of the facular knot is determined by the hydrostatic model of the solar atmosphere (Avrett & Loeser, 2008), in which photosphere is defined as that having the following parameters: $T(0) = 6583 K$, $P(0) = 1.228 \times 10^5 \, dyn \times cm^{-2}$, $\rho(0) = 2.87 \times 10^{-7} \, g \times cm^{-3}$. The layer with T = 5800 K, which is usually considered as photosphere, lies 50 km higher in this model.

### 4.1. Facular node temperature profiles at the photospheric level.

The analytical methods developed by us (Solov'ev & Kirichek, 2016; 2019) make it possible to calculate the pressure, density, temperature and velocity of plasma flow at each point of the studied magnetic configuration using a given magnetic field structure.
The pressure and density of plasma in the magnetic flux tube of a facular knot are given as follows:

$$P(r,z) = P_{ex}(z) + \frac{B_0^2}{8\pi}\left[b_r^2 + 2\int_\infty^r b_z \frac{\partial b_r}{\partial z} dr\right] - \frac{B^2(r,\varphi,z)}{8\pi} + \frac{B_{ex}^2}{8\pi}, \qquad (17)$$

$$\rho(r,z) = \rho_{ex}(z) + \frac{B_0^2}{8\pi}\frac{1}{g}\left[2b_r \frac{\partial b_z}{\partial r} - \frac{\partial}{\partial z}\left(b_r^2 - b_z^2 + 2\int_\infty^r b_z \frac{\partial b_r}{\partial z} dr\right)\right]. \qquad (18)$$

Here $P_{ex}(z), \rho_{ex,} B_{ex}^2$ are the parameters of the external environment. The value $B_{ex}$, characterizing the general magnetic field of the Sun can be neglected at the level of the photosphere. Note that the density distribution does not depend on the angle. Substituting expressions (13), (15) into (17), (18), we get:

$$P(r,z,\varphi) = $$
$$= P_{ex}(z) + \frac{B_{ex}^2}{8\pi} - \frac{B_0^2 Z^2}{8\pi}\left[J_0^2\left(1 - \frac{Z''}{Z}\right) + \left(J_0^2 + \left(\frac{Z'}{Z}\right)^2 \cdot J_1^2\right) \cdot f(r,\varphi,z)\right], \quad (19)$$

$$\rho(r,z) = \rho_{ex}(z) + \frac{B_0^2 k}{8\pi g} 2ZZ' \cdot \left[\left(1 - \frac{Z''}{Z}\right) \cdot J_1^2 + \left(1 - \frac{Z''}{2Z} - \frac{Z'''}{2Z'}\right) \cdot J_0^2\right]. \quad (20)$$

Prime over Z means differentiation by the argument (kz). The plasma temperature in the facular knot is found from the equation of state for an ideal gas.:

$$T(r,\varphi,z) = \frac{P(r,\varphi,z)\mu}{\rho(r,z)\Re}, \quad (21)$$

where $\Re, \mu$ are universal gas constant and average molar mass of gas, respectively.

Below are the figures 9-11, depicting the temperature profiles of a facular node at the photospheric level for the field strengths $B_0 = 1500G$ and $B_0 = 1000G$.

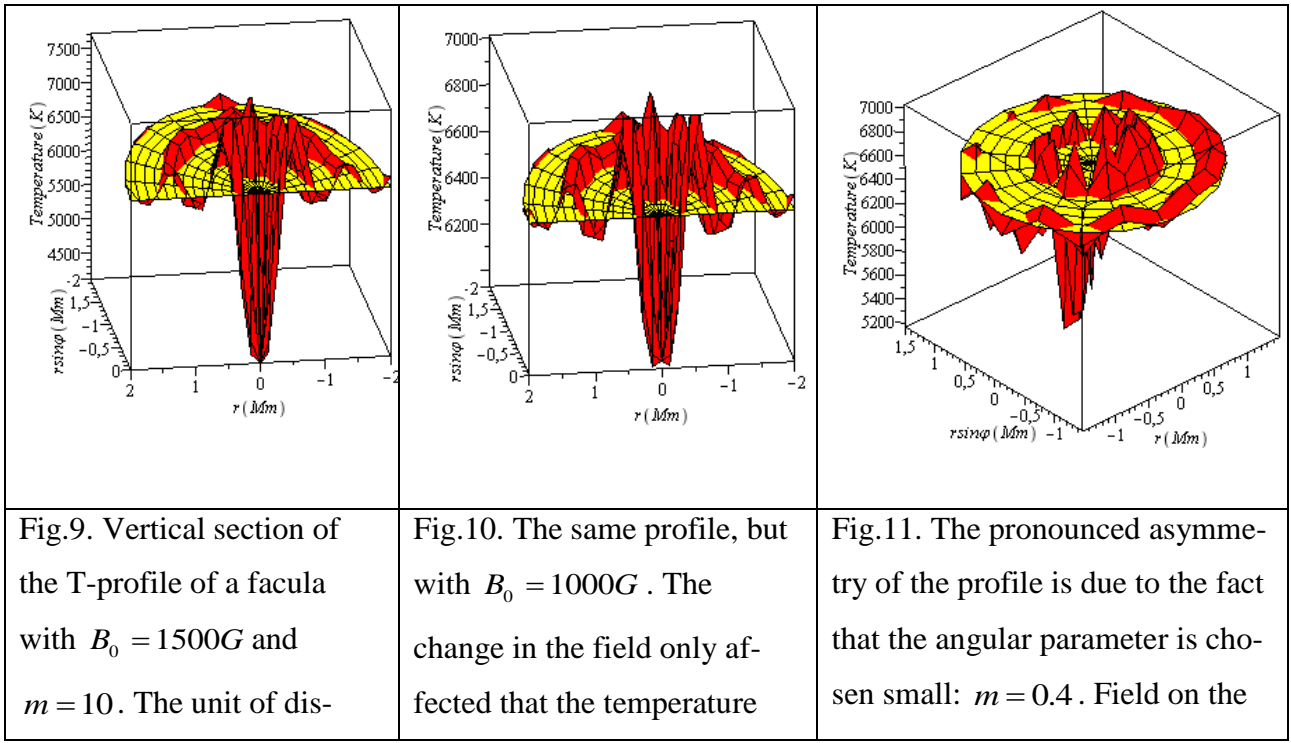

| Fig.9. Vertical section of the T-profile of a facula with $B_0 = 1500G$ and $m = 10$. The unit of dis- | Fig.10. The same profile, but with $B_0 = 1000G$. The change in the field only affected that the temperature | Fig.11. The pronounced asymmetry of the profile is due to the fact that the angular parameter is chosen small: $m = 0.4$. Field on the |

| tance along the axes is 250 km. The yellow plane is the photospheric level of T. | scale was reduced. | axis - $B_0 = 1000G$. |
|---|---|---|

As can be seen, a characteristic feature of the facular knots at the level of the photosphere is the presence of a central dip, which is similar to a micropore with a diameter of about 200 km. The greater the field strength of the magnetic field in the center of the facular knot, the greater the contrast of intensities. It follows from Figures 2-5 that the brightening surrounding the micropores, as a rule, have an asymmetric structure, which is consistent with the profile of the facula in Figure 11.

**5. Conclusions**

1. The spatial resolution of ground-based telescopes equipped with modern means of image analysis and processing reached a level where it became possible to study in detail the small-scale structure of the photosphere.

2. The continuous grid of dark intergranular spaces or lanes is an independent and extremely important element of the photospheric fine structure. Its characteristic width, like the diameter of the thinnest magnetic flux tube in the solar photosphere, is about 25 km. The resolution of modern telescopes with adaptive optics is still insufficient for a detailed study of the intergranular lanes.

3. Micropores with a diameter of 200-400 km, being the core of facular nodes, ensure their stability and thereby significantly prolong their lifetimes - from several hours to days.

4. As the simulation shows (Fig.9,10), the greater the magnetic field strength in the facular knot, the greater the central temperature dip. The direct dependence of the darkening of the center of the facula on the magnetic field strength was recently established according to the results of SDO data processing in Riehokainen et al., (2019).

**Acknowledgements.** We are grateful to the teams of Goode Solar Telescope (Big Bear Solar Observatory) and Swedish 1-meter Solar Telescope (SST) for allowing us to use the observational data.
This work was supported by the Russian National Science Foundation (project 15-12-20001), the Russian Foundation for Basic Research (project 18-02-00168) and the Program KP19-270..